\documentclass[prl,twocolumn,showpacs,superscriptaddress]{revtex4}
\usepackage{multirow,eurosym,amssymb,amsfonts,amsmath,setspace,graphicx,color,bm,float,verbatim}

\def\be{\begin{equation}}
\def\ee{\end{equation}}

\def\bsplit{\begin{split}}
\def\nsplit{\end{split}}

\begin{document}

\title{Electromagnetically-induced transparency
with amplification \\ in superconducting circuits}
\date{\today}

\author{Jaewoo Joo}
\affiliation{Institute for Quantum Information Science, University of Calgary, Alberta T2N 1N4, Canada}

\author{J\'er\^ome Bourassa}
\affiliation{D\'epartement de Physique, Universit\'e de Sherbrooke, Sherbrooke, Qu\'ebec J1K 2R1, Canada}

\author{Alexandre Blais}
\affiliation{D\'epartement de Physique, Universit\'e de Sherbrooke, Sherbrooke, Qu\'ebec J1K 2R1, Canada}

\author{Barry C. Sanders}
\affiliation{Institute for Quantum Information Science, University of Calgary, Alberta T2N 1N4, Canada}

\begin{abstract}
    We show that electromagnetically-induced transparency and lasing without inversion are simultaneously
    achieved for microwave fields by using a fluxonium superconducting circuit.
    As a result of the $\Delta$ energy-level structure of this artificial three-level atom,
    we find the surprising phenomenon that
    the electromagnetically-induced transparency window in the frequency domain is sandwiched between
    absorption on one side and amplification on the other side.
\end{abstract}


\maketitle

\begin{figure}[t]
\includegraphics[width=0.75 \columnwidth]{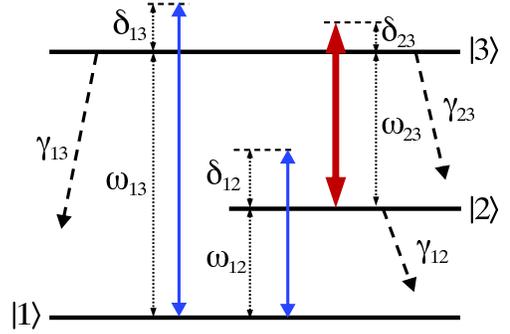} \caption{
    (Color Online)
    A $\Delta$3LA system with driving fields indicated by solid arrows (blue and red), decays by dashed lines and frequency differences between levels by dotted lines.
    }
\label{fig:Delta}
\end{figure}

Electromagnetically-induced transparency (EIT) exploits atomic coherence to enable optically-controlled transparency within an absorption line as well as extreme slowing of light~\cite{FIM05}.
EIT is realized by strong driving of one transition in a three-level atom (3LA) depicted in Fig.~\ref{fig:Delta}, which induces a transparency window with bandwidth equal to the effective splitting of the upper energy level. A 3LA also yields a distinct phenomenon known as lasing without inversion (LWI)~\cite{H89}, which corresponds to negative absorption (i.e.\ amplification) despite the lower-level population exceeding the upper-level population.
LWI arises through two interfering excitation pathways thereby suppressing absorption.
Here we show that EIT and LWI can be realized \emph{simultaneously}
(as EIT with amplification, or EITA)
via a 3LA with all three inter-level transitions being driven in a $\Delta$ configuration (a $\Delta$3LA).
Furthermore we show that this new phenomenon can be realized with
flux~\cite{MOL+99} or fluxonium~\cite{MKDG+09} artificial atoms,
which are solid-state realizations of $\Delta$3LAs~\cite{LYW+05}.
Our theory of EITA predicts asymmetric EIT peaks, commensurate with experimental observations
of an anomalous asymmetry of EIT peaks for Rb $\Delta$3LAs~\cite{LSV+09}.

We construct a theory of EITA and build on recent advances with Josephson-junction-based
artificial atoms to show how
EITA can be achieved and what its experimental signature will be.
There are subtleties though in transferring optical atomic
experiments to the superconducting circuit domain. In particular
the superconducting circuit employs microwave fields that
propagate in one dimension in contrast to three-dimensional field
propagation and optical frequencies in the atomic experiment.
Therefore absorption and transmission spectroscopy translate to
reflected and transmitted fields. Also optical experiments employ
a large number of 3LAs whereas the superconducting circuit case
needs just one 3LA. With EIT and the Autler-Townes splitting
having been displayed in experiments, artificial $\Delta$3LAs
built with superconducting junctions appear to be good candidates for
observing EITA~\cite{DMOO+06,BFB+09,SLC+09,ILN+09}.

The $\Delta$3LA depicted in Fig.~\ref{fig:Delta} has three energy levels~$\left|i\right\rangle$ with  frequency differences~$\omega_{ij}$ and decay rates  $\gamma_{ij}$ between levels $\left|i\right\rangle$ and $\left|j\right\rangle$ for $i,j\in\{ 1,2,3\}$.
The $\left|i\right\rangle\leftrightarrow\left|j\right\rangle$ transition is driven by a coherent electromagnetic field
with electric field $\bm{E}_{ij}$ and detuning~$\delta_{ij}$ from the $\left|i\right\rangle\leftrightarrow\left|j\right\rangle$ transition with electric dipole vector~$\bm{d}_{ij}$. The corresponding (complex) Rabi frequency is $\Omega_{ij} =\bm{d}_{ij}\cdot\bm{E}_{ij}$ ($\hbar\equiv 1$).
Away from the flux degeneracy point, selection rules do not apply to this one-dimensional superconducing circuit so all dipole transitions can be driven by applying a trichromatic microwave field tuned
near each of the $\omega_{ij}$~\cite{LYW+05}.

For $\hat{\sigma}_{ij}:=\left|i\right\rangle\left\langle j\right|$, the system Hamiltonian is
\begin{equation}
     \hat{H}= \sum^3_{i=1}\omega_i\hat{\sigma}_{ii}
        -\frac{1}{2}\sum_{i>j}\left(\Omega_{ij} \text{e}^{- i \left( \omega_{ij} +\delta_{ij}\right)t}\hat{\sigma}_{ij}
    + \text{hc}\right),
\label{eq:Hsys}
\end{equation}
with hc denoting the Hermitian conjugate.
For a rotating frame and taking $\delta_{12}=\delta_{13}-\delta_{23}$, Eq.~(\ref{eq:Hsys}) is replaced by
\begin{equation}
\label{DeltaHamil01}
\hat{H}_\text{int}
    = - \sum_{i=2}^{3} \delta_{1i}\hat{\sigma}_{ii}
        -\frac{1}{2}  \sum_{i>j} \left(  \Omega_{ij}\hat{\sigma}_{ij}
        + \text{hc}\right).
\end{equation}
Energy relaxation and dephasing caused by coupling to uncontrolled degrees of freedom are described
by a Lindblad-type master equation
\begin{align}
\label{eq:Lindblad}
\dot{\rho}
    &=- i [\hat{H}_\text{int},\rho] + \sum_{i<j} \gamma_{ij} \mathcal{D}[\sigma_{ij}]\rho
    + \sum_{i=2}^3 \gamma_{\phi i} \mathcal{D}[\sigma_{ii}]\rho
        \nonumber   \\
    &=: \mathcal L\rho
\end{align}
for $\mathcal D[c] \bullet := c \bullet \hat{c}^\dag-\{\hat{c}^\dag c,\bullet \}/2$.
Here $\gamma_{\phi i}$ is a pure dephasing rate for level~$\left|i\right\rangle$,
which should be negligible for flux $\Delta$3LAs at the flux degeneracy point~\cite{BCB+04} and for fluxonium $\Delta$3LAs in a wider range of flux around this degeneracy point~\cite{MKDG+09}.

\begin{figure}[t]
\includegraphics[width=0.8 \columnwidth]{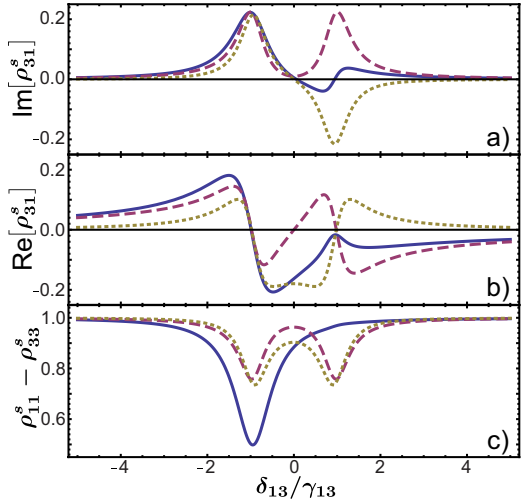}
\caption{
    (Color online)
    a)~Absorption, b)~dispersion and c)~population inversion vs detuning $\delta_{13}/\gamma_{13}$ for
    EIT (red dashed lines), LWI (yellow dotted lines) and EITA (blue solid lines) for
    $10\gamma_{12}=\gamma_{13}$, $\gamma_{23}=\gamma_{12}$, $\delta_{23}=0$,
    and $5\Omega_{13}=5\Omega_{12}=\gamma_{13}$ and $\Omega_{23}=\gamma_{13}$.
    }
\label{fig:EITAIm}
\end{figure}

For EIT, a strong pump field $\Omega_{23}\gg\Omega_{13}>0$ causes
Autler-Townes splitting of level $\left|3\right\rangle$ yielding
two absorption peaks at
$\delta_{13}/\gamma_{13}=\pm\left|\Omega_{23}\right|/2\Gamma_3$
for~$\Gamma_3=(\gamma_{13}+\gamma_{23}+\gamma_{\phi3})/2$  with a transparency window centered at
$\delta_{13}=0$ and full-width at half-maximum
$\text{FWHM}=\gamma_{12}+\gamma_{\phi2}+|\Omega_{23}|^2/2\Gamma_3$~\cite{FVA+06}.
Optical dispersion and absorption are quantified, respectively,
by the real and imaginary parts of the first-order
susceptibility~\cite{ILN+09} $\chi^{(1)} \propto |d_{13}|^2
\rho^\text{s}_{31}/\Omega_{13}$ with $ \rho^\text{s}_{ij}=\langle
i|\rho^\text{s}|j\rangle$ the steady-state solution of the master
equation. Hence dispersion and absorption are proportional to
Re$\left[\rho^\text{s}_{31}\right]$ and
Im$\left[\rho^\text{s}_{31}\right]$, shown in
Figs.~\ref{fig:EITAIm}(a,b) for EITA, EIT ($\Omega_{12}=0$) and
LWI ($\Omega_{13}=0$). As expected, these dispersion and
absorption curves are related by the Kramers-Kronig relation.

The EIT absorption curve exhibits a transparency window between
two Autler--Townes peaks, and the linear dispersion curve in
Fig.~\ref{fig:EITAIm}(b) indicates that the group velocity is
constant in this window. The LWI absorption curve shows the
characteristic transparency at resonance with absorption in the
red-detuned (left) region and amplification (or negative
absorption) in the blue-detuned (right) region. EITA exhibits the
transparency window characteristic of EIT but with the LWI
feature that the window is bounded by an absorption and an
amplification peak rather than by two Autler-Townes absorption
peaks. Fig.~\ref{fig:EITAIm}(c) confirms that population
inversion $\rho^\text{s}_{11} - \rho^\text{s}_{33}$ is always
positive for EIT, LWI and EITA so amplification is not due to
population inversion.

\begin{figure}[t]
\centering
\includegraphics[width=0.8 \columnwidth]{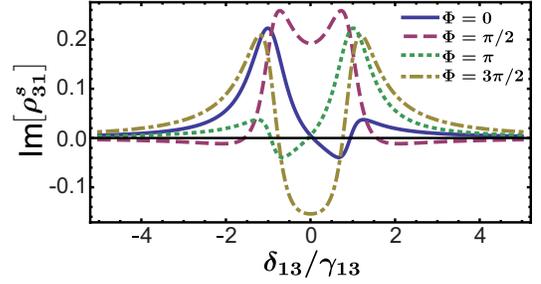}
\caption{
    (Color online)
    Absorption vs detuning as obtained from the steady-state of Eq.~(\ref{eq:Lindblad}) for $\Phi$
    equal to $0$ (solid blue line), $\pi/2$ (dashed red line),
    $\pi$ (dotted green line) and $3\pi/2$ (dash-dotted yellow line).
    Parameters are the same as in Fig.~\ref{fig:EITAIm}.
    }
\label{fig:Rho13Phase}
\end{figure}

In fact, EITA is not a simple combination of EIT and LWI as coherence between each pair of levels adds to the richness of the phenomenon. Due to inter-level coherence, controlling the relative phase of (at least) one field with respect to the other two affects whether amplification is in the red- or blue-detuned region or even whether there is amplification at all, as depicted in Fig.~\ref{fig:Rho13Phase}.
This control becomes evident by taking $\rho^{s}_{23} \approx 0$~\cite{footnote:lead}:
\begin{equation}
\begin{split}
\rho^{s}_{31}
    =& -{\rm e}^{ - i \phi_{13}} \bigg[ 2 i\Omega_{13}
    \left( \rho^{s}_{11} - \rho^{s}_{33} \right)
    \left( i\delta_{13} - \gamma_{12}/ 2 \right) \\
    & + \Omega_{23} \Omega_{12}
    {\rm e}^{- i \left( \phi_{12} + \phi_{23}-\phi_{13}\right)}
    \left( \rho^{s}_{11} - \rho^{s}_{22} \right) \bigg] / F,
    \label{eq:RhoDelta02}
\end{split}
\end{equation}
where $F= 4\left( i \delta_{13} - \Gamma_{3} \right) \left( i \delta_{13} - { \gamma_{12}/ 2} \right) + \Omega_{23}^2$.
For $\Phi:=  \phi_{12}+\phi_{23}-\phi_{13}$
we observe that the absorption curve of Fig.~\ref{fig:EITAIm}(a)
is recovered for $\Phi=0$. EITA is replaced by ordinary absorption
for $\Phi=\pi/2$ with $\gamma_{12} \ll \gamma_{13}$,
and the mirror image of the $\Phi=0$ absorption curve occurs
for $\Phi=\pi$. For $\Phi=3\pi/2$, the absorption curve
corresponds to an EIT profile but with the transparency window
replaced by an amplification window accompanied by a linear dispersion profile (not shown) so group
velocity is constant for this window.

\begin{figure}[t]
\centering
\includegraphics[width=0.9 \columnwidth]{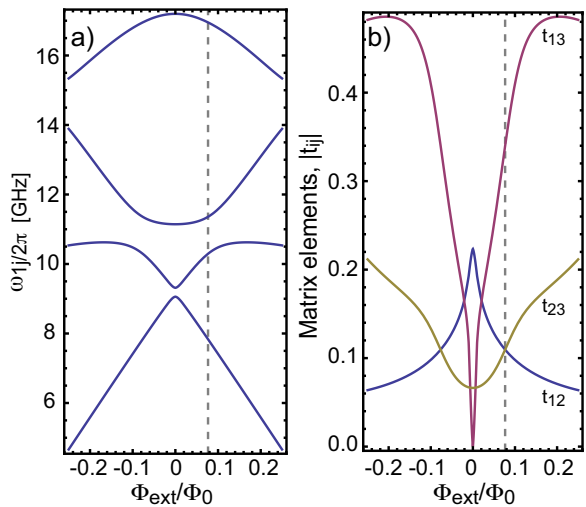}
\caption{
    (Color online)
    a) Transition frequencies $\omega_{i0}/2\pi$ and b) coupling matrix elements $\left|t_{ij}\right|$
    for charge coupling in the fluxonium $\Delta$3LA between states $\left|i\right\rangle$
    and $\left|j\right\rangle$ as a function
    of the external flux $\Phi_\mathrm{ext}/\Phi_0$, with $\Phi_{0}$ being the flux quantum.
    Parameters are from Ref.~\cite{MKDG+09}.
    We suggest biasing the artificial 3LA at $\Phi_\mathrm{ext}/\Phi_0 = 0.08$, indicated
    by a dashed vertical line, where $t_{12}=t_{23}$, optimal for observation of EITA.
    }
\label{fig:MatrixElements}
\end{figure}

Our theory of a $\Delta$3LA is applicable to a recent EIT experiment with Rb atomic gas \cite{LSV+09},
which exhibited both transmission enhancement and asymmetry between the red- and blue-detuned
transmission peaks.
Their theory explains transmission enhancement but not the observed peak asymmetry.
As the lower two levels of their Rb 3LA is driven by a microwave field, their system is the $\Delta$3LA discussed here,
and our theory predicts  the observed peak asymmetry although, of course, a quantitative analysis is required
to see how much of the asymmetry is due to $\Delta$ electronic structure effects as opposed to other reasons.
Although our theory also predicts negative absorption (amplification),
inhomogeneous broadening and absorption in the gas cell could obscure the amplification signature.
An advantage of our proposal to study EITA with superconducting artificial atoms coupled to one-dimensional transmission lines is that EITA can be investigated in a controlled way without some of the
complications that arise for gases.

Flux~\cite{MOL+99} and fluxonium~\cite{MKDG+09} 3LAs closely approximate
$\Delta$3LAs away from flux degeneracy~\cite{LYW+05}, hence are natural candidates for realizing EITA.
Fig.~\ref{fig:MatrixElements}(a) shows the energy levels structure of the fluxonium 3LA
and Fig.~\ref{fig:MatrixElements}(b) the corresponding transition matrix elements $|t_{ij}|$, both as a function of the externally-applied flux $\Phi_\mathrm{ext}$.
Away from $\Phi_\mathrm{ext}/\Phi_0=0$ and $0.5$, all three matrix elements have comparable values such that the fluxonium can be used as a $\Delta$-system. Flux 3LAs have similar magnetic flux transition matrix elements~\cite{LYW+05} but are more sensitive to flux noise and tend to have the state $\left|3\right\rangle$ high in energy.
Under these considerations, the fluxonium appears to be a promising candidate for the observation of the effect studied here.

\begin{figure}[t]
\centering \includegraphics[width=0.8\columnwidth]{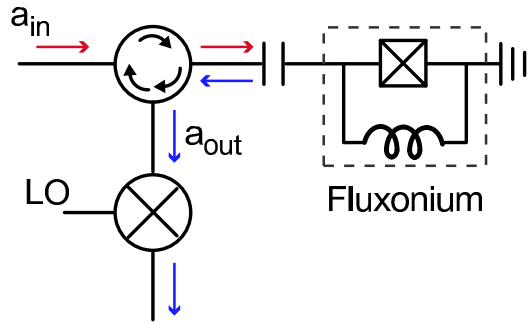}
\caption{
    (Color online)
    A fluxonium 3LA, made from the parallel combination of a Josephson junction and a large inductance, is capacitively coupled to the end of a semi-infinite transmission line and measured in reflection. A circulator is used to separate the input and the output fields. The output field is amplified (not shown) and mixed with a local oscillator (LO) to realize a homodyne measurement.
    }
\label{fig:homodyne}
\end{figure}

EITA can be probed by connecting either 3LA to a transmission line supporting traveling modes~\cite{AZA+10}. The absorption and dispersion profiles can be measured both in transmission and reflection, and a possible setup for homodyne measurement of the reflected signal is illustrated in Fig.~\ref{fig:homodyne}. To determine how the reflected signal contains information about $\rho_{31}$, we use
input-output theory~\cite{GC85}.
In the Markov approximation and focusing on the signals centered about the probe ($a$),
pump ($b$) and control ($c$) frequencies, the transmission-line free Hamiltonian is
\begin{equation}
    \hat{H}_\mathrm{TL} = \sum_{\hat{o}\in\{\hat{a},\hat{b},\hat{c}\}}
    \int_0^\infty \text{d}\omega\,\hat{o}^\dag(\omega)\hat{o}(\omega)
\end{equation}
with the microwave field annihilation operators $\hat{o}(\omega)$ satisfying $[\hat{o}(\omega),\hat{o}'^\dag(\omega')]=\delta_{\hat{o},\hat{o}'}\delta(\omega-\omega')$.

Treating the transmission-line mode as three commuting quasi-monochromatic modes
is valid if separation between the transitions frequencies greatly exceeds the linewidths.
In this approximation, the $\Delta$3LA-transmission line interaction Hamiltonian is
\begin{align}
    \hat{H}_\text{int}
        =&  i  \int_{-\infty}^\infty \text{d} \omega \Big[ \sqrt{\frac{\gamma_{13}}{2\pi}}
            \hat{a}^\dag(\omega)\hat{\sigma}_{13}
            +\sqrt{\frac{\gamma_{23}}{2\pi}} \hat{b}^\dag(\omega)\hat{\sigma}_{23}\nonumber \\
        &+\sqrt{\frac{\gamma_{12}}{2\pi}} \hat{c}^\dag(\omega)\hat{\sigma}_{12}- \text{hc}\Big].
\end{align}
Using input-output theory, the output field operator centered at the probe frequency is
\begin{equation}
    \hat{a}_\text{out}(t) = \hat{a}_\text{in} (t)  +  \sqrt{\gamma_{13}}\hat{\sigma}_{13}(t),
\end{equation}
with $\hat{a}_\text{in} (t)$ the annihilation operator for the input field centered at the probe frequency.
With the homodyne setup illustrated in Fig.~\ref{fig:homodyne} effectively measuring  $\langle \hat{a}_\text{out}(t)\rangle = \langle \hat{a}_\text{in} (t)\rangle+\sqrt{\gamma_{13}} \rho_{31}(t)$, access to the dispersion and absorption profiles is straightforward.

We propose to bias the fluxonium at $\Phi_\mathrm{ext}/\Phi_0 =
0.08$, indicated by a dashed vertical line in
Fig.~\ref{fig:MatrixElements}, where $t_{12}=t_{23}$ as this
choice is optimal for observation of EITA.
Contrary to the case of resonators~\cite{BHW+04}, coupling to the transmission line
traveling modes exposes the $\Delta$3LA to environmental vacuum
fluctuations of the voltage at the $\Delta$3LA transition frequencies,  thereby enhacing
spontaneous decay. As an estimate for the relaxation time in this case, we use Astafiev
et al.'s results where a flux qubit was
coupled to a transmission line with $\gamma_{12}/2\pi =11$~MHz at zero flux~\cite{AZA+10}.
Assuming white noise, the decay rates at $\Phi_\mathrm{ext}/\Phi_0 = 0.08$
can be estimated using the matrix elements of Fig.~\ref{fig:MatrixElements}, yielding $\gamma_{13}/2\pi = 25$~MHz, $\gamma_{12}/2\pi = 2.6$~MHz and $\gamma_{23}/2\pi = 2.6$~MHz.   Figs.~\ref{fig:EITAIm} and \ref{fig:Rho13Phase} have been obtained using these values, showing that EITA with superconducting $\Delta$3LA should be possible with
current experimental parameters.

Another approach to probing EITA with superconducting circuits is by quantum state tomography where the density matrix is fully reconstructed.  This can be done, for example, by coupling the $\Delta$3LA to a resonator rather than a transmission line~\cite{FML+09},
and strong coupling of a flux $\Delta$3LA to a resonator has been studied~\cite{BGA+09,AAN+08,NDH+10}.
An advantage of this approach is that the resonator will shield the $\Delta$3LA from noise away from the resonator frequency, thereby decreasing significantly the decay rates.

In summary, we have developed the theory of EITA which
shows a EIT window sandwiched between an absorption line and an
amplification line in a superconducting $\Delta$3LA system. The EITA absorption and dispersion
profiles can be controlled by the phase of one of the three microwaves
applied to the superconducting atom.
We suggest a homodyne measurement scheme for a direct observation of the EITA absorption and dispersion profiles of the probe field where a fluxonium artificial atom is coupled to a one-dimensional transmission line.
EITA is exciting as a surprising combination of electromagnetically-induced transparency in a single system,
and superconducting artificial atom realizations will enable controlled study of this phenomenon.
EITA could be useful for superconducting circuits by enabling slowing and storage of microwave fields,
and the amplification effect could be useful for partially offsetting absorption.

\begin{acknowledgments}
This project is supported by FQRNT, NSERC, \emph{i}CORE, and QuantumWorks.
We thank S. Filipp for discussions about the phase control of multiple microwave drives.
AB is partially supported by the Alfred P.\ Sloan Foundation.
AB is a CIFAR Scholar, and BCS is a CIFAR Fellow.

\end{acknowledgments}


\begin{thebibliography}{11}
\bibitem{footnote:lead}
With the parameters of Fig.~\ref{fig:EITAIm}, we find $\text{Max}[|\rho^{s}_{23}|] \approx 0.15$ hence the approximation $\rho^s_{23} \approx 0$ does not lead to a good quantitive analysis.
\end{thebibliography}
\end{document}